\documentclass{emulateapj}
 \usepackage{natbib}

\def\phs{ph~cm$^{-2}$~s$^{-1}$}

\slugcomment{Accepted for publication in Astrophysical Journal Letters}

  \shortauthors{Roques et al.}
 \shorttitle{First observations of the Hard X-ray emission of V404 Cyg 2015 outburst with \textit{INTEGRAL}}

\begin{document}

\title{FIRST \textit{INTEGRAL}\footnotemark[1] OBSERVATIONS OF V404 CYGNI DURING THE 2015 OUTBURST: SPECTRAL BEHAVIOR IN THE 20 - 650 KEV ENERGY RANGE
}
\footnotetext[1]{Based on observations with \textit{INTEGRAL}, an ESA project with instruments and science data centre funded by ESA member states
 (especially the PI countries: Denmark, France, Germany, Italy, Spain, and Switzerland), Czech Republic and Poland with participation of
 Russia and USA.}

\author{Jean-Pierre~Roques\altaffilmark{1},Elisabeth~Jourdain\altaffilmark{1}\\
Angela Bazzano\altaffilmark{2}, Mariateresa Fiocchi\altaffilmark{2}, Lorenzo Natalucci\altaffilmark{2}\& Pietro Ubertini\altaffilmark{2}}

\affil{\textsuperscript{1}Universit\'e Toulouse; UPS-OMP; CNRS; IRAP; 9 Av. Roche, BP 44346, F-31028 Toulouse, France\\
\textsuperscript{2}Istituto di Astrofisica e Planetologia Spaziali, INAF, Via Fosso del Cavaliere 100, Roma, I-00133, Italy
}

\begin{abstract}
In June 2015, the source V404 Cygni (= GS2023+38) underwent  an extraordinary 
outburst. 
We present the results obtained during the first revolution dedicated to this 
target by the \textit{INTEGRAL} mission, and focus on the spectral behavior in the hard 
X-ray domain, using both SPI and IBIS instruments. The source exhibits extreme variability,
and reaches fluxes of several tens of Crab. However, the emission between 20 
and 650 keV can be understood in terms of two main components, varying 
on all the observable timescales, similar to what is observed in 
the persistent black hole system Cyg X-1. The low energy component (up to $\sim$ 
200 keV) presents a rather unusual shape, probably due to the intrinsic source variability.
Nonetheless, a satisfactory description is obtained with a Comptonization model, if 
 an  unusually hot population  of seed photons ($kT_0 \sim$ 7 keV) is introduced.
 Above this first component, a clear excess extending up to 
400-600 keV leads us to investigate a 
scenario where an additional (cutoff) power law could correspond to the 
contribution of the jet synchrotron emission, as proposed in Cyg X-1. 
A search for an annihilation feature did not provide any firm detection, with
 an upper limit of $2 \times 10^{-4}$ \phs\  ($2 \sigma)$
for a narrow line  centered at 511 keV, on the averaged obtained spectrum. 
\end{abstract}      


\maketitle

\section{Introduction}
The 2015 outburst of V404 Cyg will probably remain a unique event in 
astrophysical history. Identified as a nova in 1938, when only optical 
observations were available, V404 Cyg came back to the forefront in 1989 
when another nova episode was observed simultaneously in optical and in the soft X-rays 
and hard X-rays domains by the GINGA satellite (from where its second name  comes
GS2023+338, \citealt{mak89}) and the Roentgen instruments onboard the Kvant module \citep{sun91}. It 
presented  huge variability, and reached several Crab flux levels, in soft X-rays as 
well as in hard X-rays, becoming the brightest source ever observed in these 
energy ranges. The parameters have been determined, with the
system  located  at 2.39 $\pm$ 0.14 kpc \citep{MJ09} and containing a $\sim$ 10 ${M}_{\rm \odot}$  black hole, with an orbital period around
 6.5 days \citep{Casares92}.
The 2015 episode started with similar burst-like activity that was first detected by 
Swift/BAT \citep{GCN17929} and Fermi/GBM \citep{GCN17932}, and soon
after was observed at all wavelengths from radio \citep{radioatel} to hard X-rays and reached exceptional levels of 
luminosity, exceeding all expectations (up to 50 Crab in hard X-rays, \citealt{JeromeV404}). 

Among the global monitoring of the source, \textit{INTEGRAL}, with its two main instruments 
SPI and IBIS, played a major role in the hard X-ray/soft $\gamma$-ray studies. 
\section{Instrument, observations and data analysis}\label{instru}
We analyzed the first public observation from 2015 June 17 to June 20 (revolution 1554) for a total 
useful duration of 150 ks. The SPI analysis is based on consolidated data
and algorithms developed at IRAP
\footnote{An open SPI Data Analysis Interface (\emph{SPIDAI}) 
allows one to perform the SPI data analysis with  the same  
tools as those used in this paper. See the dedicated webpage http://sigma-2.cesr.fr/INTEGRAL/spidai}.
Considering the source exceptional flux level and variability, it is worth mentioning that the SPI
data are not affected by pile-up or TM saturation.  
Only a limited number of data packets are missing corresponding to a few seconds of data, with an energy independent effect.
The IBIS data for this observation are
near real-time data processed using the latest release of
the \textit{INTEGRAL} Offline Scientific Analysis (OSA version 10.1). 
The IBIS and SPI results have been compared, and show that the source
fluxes are in good agreement, even if, at high flux
levels, IBIS/ISGRI spectra appear systematically harder than the SPI ones
 (See analysis by \citealt{Natalucci et al. 2015}). The
reason for this is under investigation, and probably related to a
combination of source variability and telemetry saturation. \\
The description of the instruments and performance can 
be found in  \citet{Vedrenne et al. 2003} and \citet{Roques03} for SPI, 
and \citet{Ubertini et al. 2003} for IBIS.
\section{Broad band and spectral evolution}\label{lightcurve}
\begin{figure*}
\plotone{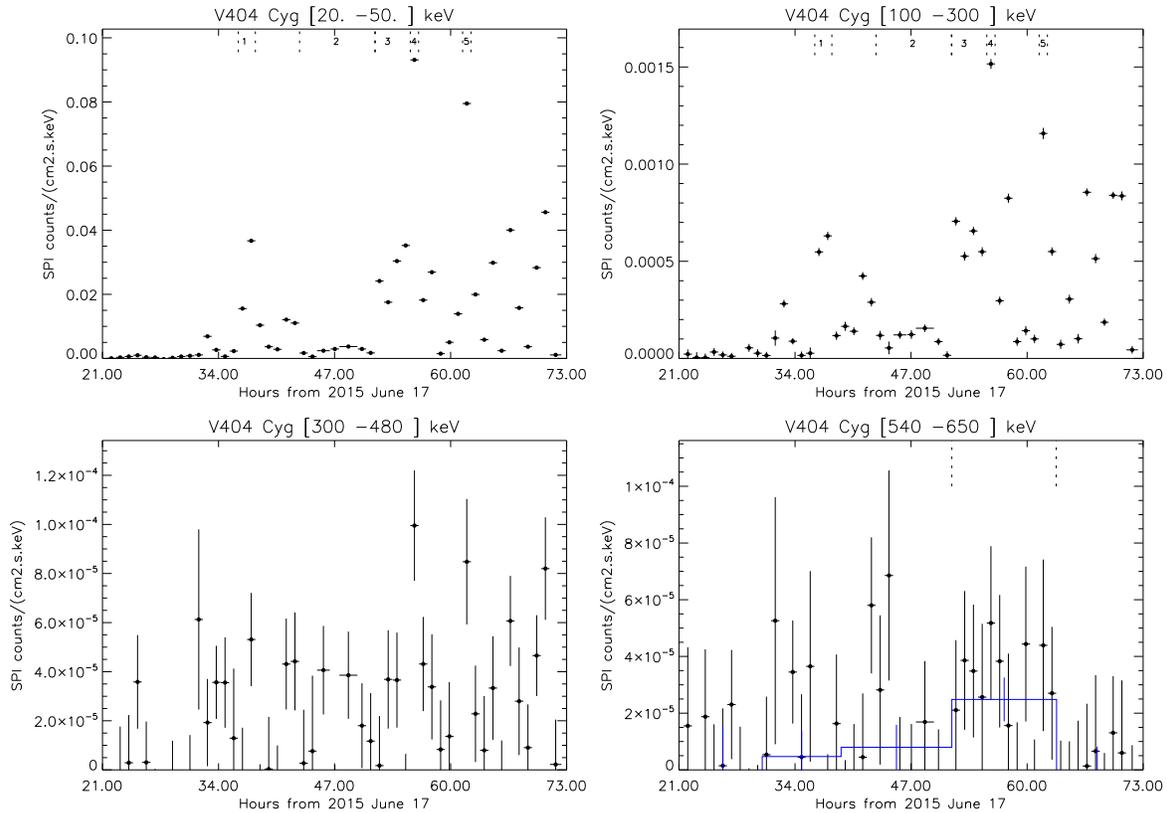} 
\caption{SPI light curves of V404 Cyg from June, 17 to 20,
on the scw timescale ($\sim$ 1 h). In the last panel, the blue histogram corresponds to a $\sim$ 10 h timescale}
\label{fig:cl.1swc}
\end{figure*}

The source variability at all wavelengths and on all timescales is striking.
However, the emission above $\sim$ 200 keV contains key information about the  population(s) 
present and energy transfer but requires longer  integration time. 
The present analysis is based on the science window (scw) timescale, a time interval lasting 
approximatively 3.4 ks, and corresponding to a stable pointing direction. 
Fig. \ref{fig:cl.1swc} displays the source evolution over 4 broad bands covering the 20-650 keV domain,  
and demonstrates that, on this $\sim$ hour timescale, the source variability is significant  up to $\sim$ 500 keV,
and appears energy dependent.

\begin{figure}
 \plotone{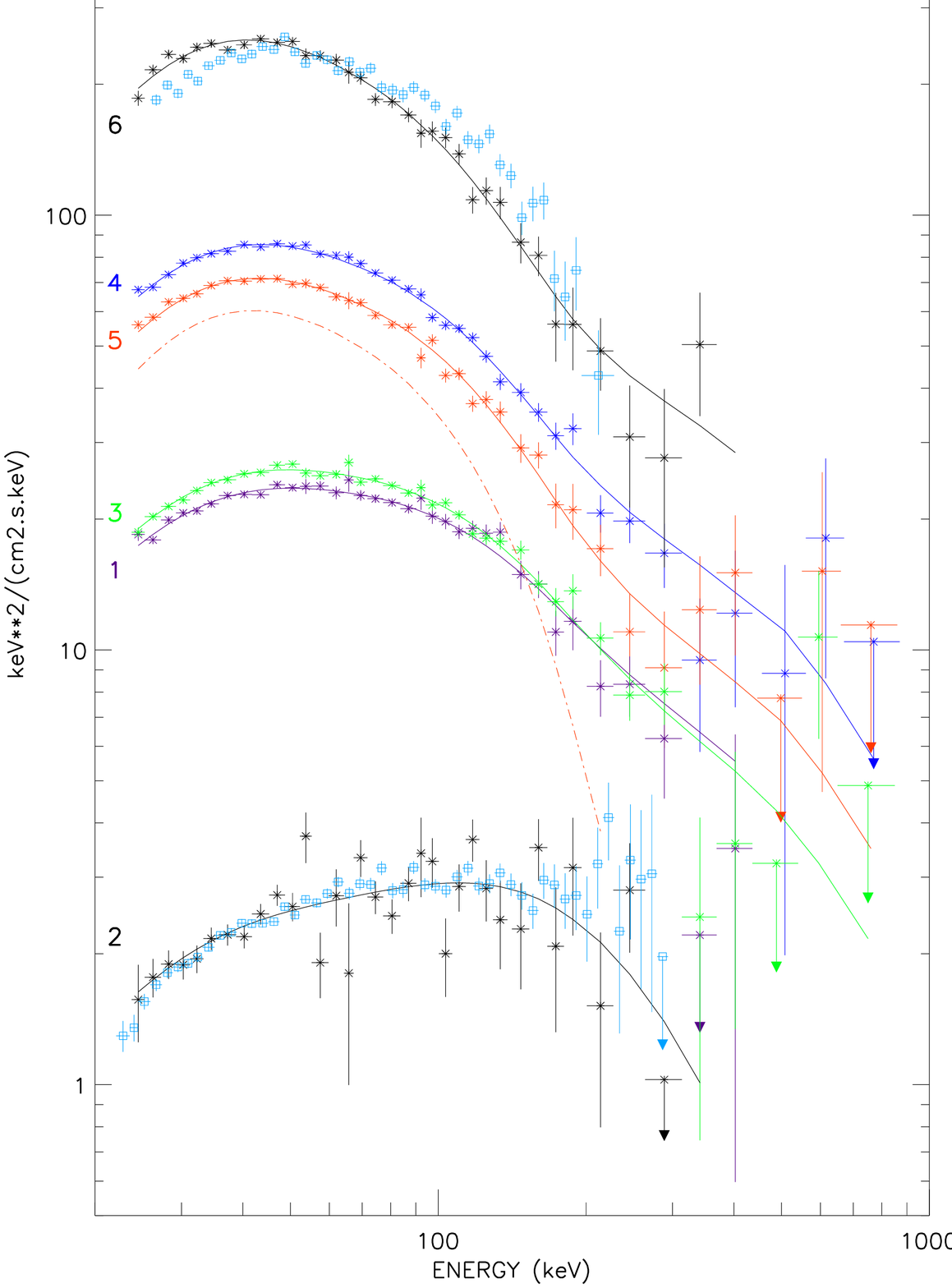}
 \caption{Evolution of the spectral shape of V404 Cyg, on the 1-4 hours (labels 1 to 5)
and 100 s (label 6) timescales. Crosses correspond to SPI data, open squares to IBIS/ISGRI data (for labels 2 and 6 only). Solid lines are the best common fit models, given in Tab.\ref{tab:fitindiv} .
See text for the period definition.}
\label{fig:6spectres}
\end{figure}

In order to quantify the spectral evolution,  the observed emission 
is described with a common model, both physically reasonable and as simple as possible.
An averaged spectrum has been built 
and  fitted with the {\sc xspec} v12.8.2 tools \citep{xspec}. Starting from a Comptonization model , we
note the emergence of a high energy component around 100 keV. This component is reminiscent of the hard tail observed in several X-ray binary systems, transient sources (e. g. Nova Persei=GROJ0422+32, \citealt{NP}) or persistent sources (e. g. Cyg X-1).
Continuing the analogy with the well-studied source Cyg X-1, we add to our model the same second component (cutoff powerlaw )
used to account for the high energy tail in the hard state.
This simple model failed to take into account the actual curvature in the low energy part, and we have tested a scenario with
a reflecting and/or absorbing medium. The required values imply very specific geometry or a strong absorption of photons at lower energy. As a second option, we relaxed the constraint on the temperature of seed photons ($kT_{0}$). Surprisingly, a temperature around 6-7 keV 
perfectly reproduces the data.  \\
Finally, we have chosen to describe the spectral emission of V404 Cyg with the following model:  a Comptonization component (Comptt), with two free parameters (kT and $\tau$, temperature and optical depth of the Comptonizing electron population, with $kT_{0}$ 
 fixed  to 6.5 or 7 keV, see Table \ref{tab:fitindiv}) and a cutoff power law, to account for the high energy part. We fix its photon index to 1.6  and start with a cutoff energy default value of 300 keV.\\
From Fig. \ref{fig:cl.1swc}, a few periods have ben chosen to illustrate the spectral evolution of V404 Cyg:\\
(1)First small peak (June 18, 12:11-14:07 UT; 6 ks)\\
(2)'Off-flare' state (June 18, 19:02-June 19, 03:29 UT; 25 ks)\\
(3)Plateau (June 19, 03:31-07:25 UT; 11.5 ks)\\
(4)First maximum peak (June 19, 07:26-08:23 UT; 3 ks)\\
(5)Second maximum peak (June 19, 13:19-14:15 UT; 3 ks)\\
The corresponding spectra are compared in Fig.\ref{fig:6spectres}. Tab.\ref{tab:fitindiv} gives the best fit parameter values  
when they are fit individually, and the parameters obtained for the averaged spectrum 
for comparison. For the low flux periods ([1] and [2]), the cutoff powerlaw component is not required, probably
 due to the low signal to noise ratio. Note that these periods are described by a hotter and thiner Comptonizing medium, while the temperature  decreases and 
the optical depth increases when the source's flux is high. 
In a second step, to overcome the error bars and degeneracy between kT and $\tau$,  we have also fit all the spectra together, 
aiming to extract  more information with a minimal number of free  parameters. We thus started 
 our fitting procedure by imposing the same parameter values (except normalizations) to all considered spectra, and freed successively, one by one, those required to improve significantly the global $\chi2$. Finally (see Table\ref{tab:fitindiv}), a global reasonable description is  obtained with $kT_{0}$  fixed to 7.0 keV and photon index to 1.6, while a common $\tau$ converges toward a value of  1.55. The cutoff energy  is fixed to 300 keV  for all spectra but the averaged one, which includes a lot of spectral variability, limiting its scientific contents.
The most striking result is that, in such a scheme, the temperature is stable (between 25 and 28 keV), except during the  period where the source seems to be less active (period [2]). The spectral emission at that time is harder, and this translates into a Comptonizing temperature of 40-50 keV.
It is worth noting that the high energy component is formally required in the high flux level spectra, and that introducing it in the
 moderate flux peak (period [1]) affects the best fit parameters in such a way that they become
 compatible with those obtained during the high flux periods (see Tab.\ref{tab:fitindiv}).
Notice also that at shorter timescales, the spectral shape evolution can be described with the same scheme. In Fig.\ref{fig:6spectres},
 spectrum [6] corresponds to the maximum in a 100 s time bin. When  this short specrum is compared to the longer term
 spectra, the global shapes appear similar, indicating that low and  high energy components evolve 
more or less  in the same way at all time scales.  
This information, together with correlations at other wavelengths,  will be crucial to 
 draw a more complete picture of the emitting regions and their intrinsic evolution.\\ 

As a supplementary study, we have considered  
a 4.5 keV wide channel centered at 511 keV in order to search for an annihilation emission.
The observed flux remains below 2 $\sigma$ for each of the 
individual scws, leading to an upper limit of 1-2 $\times 10^{-3}$ \phs.
When considering the whole revolution (150 ks), we obtain a 2~$\sigma$ upper limit of 2~$\times~10^{-4}$ \phs.
Compared to the  values given for other black hole binaries in Teegarden and Watanabe (2006), these upper limits can provide further  constraints 
for  models since they give limits in the case of very high luminosity, and on short timescales.\\

\section{discussion}
 
We are reporting here results obtained in the  hard X-ray domain (20 to $\sim$ 650 keV) from the first revolution dedicated 
to V404 Cyg by the INTEGRAL mission. While the extreme variability makes it difficult to 
get instantaneous values of the parameters of emitting region(s), we studied the source evolution
during the flaring activity to estimate values representative of the global emission.
We have shown  that  the observed emission can be described with two components. 
Several scenarios are able to account for the data, but we had to choose one to present 
the results in a way as instructive as possible. 
In this work, the first component is identified with a thermal Comptonization emission. 
The low energy curvature has required a specific attention. Three parameters can help to describe it properly: An absorbing factor, $N_{h}$,
a reflection factor, R, or the temperature of the seed photons, $kT_{0}$. However, for each of them, the fit procedure converges toward unusual values: 
$N_{h}$ of the order of a few $\times 10^{24}cm^{-2}$, in contradiction with the soft X-rays observations \citep{swiftatel}, R $\sim$ 5, which would imply a peculiar geometry or $kT_{0} \sim$ 7 $\pm 1$ keV. This may be linked to the exceptional source luminosity or reflect  the presence of another component at low energy or else another origin of the seed photons. The unceasing variability of the
 Comptonizing population parameters, in space and/or in time, may also be responsible for the observed shape, since any gradient in kT or $\tau$ modifies the resulting emission.  We give in this paper the results obtained with $kT_{0}$ = 7 keV. The reflection component
 is not included since it would add an additional free parameter which cannot be constrained. We have checked that to add a reflection factor (fixed to 1) only slightly modifies the best fit parameter values and does not affect the scientific conclusions.\\ 
For the second component, we have chosen a cutoff power law with a photon
 index of 1.6 and a cutoff energy of 300 keV, to test a scenario similar to that observed in Cyg X-1 \citep{Cygpolar}.\\
With the above hypothesis, we obtain a good description of the data and  identify  two states. 
a) A quiet phase, when the source flux remains relatively low (300 mCrab) and the spectrum hard, with a Comptonization temperature 
$\sim$~40 keV. b) An "active" phase, when the source flares,
 with huge variability and impressive amplitude changes, and a plasma temperature around 25 keV. 
The second component is not required during the quiet phase and varies significantly during the active phase.
Both components contribute to the global flux intensity,  even if their variations are 
not strictly correlated, leading to an evolving spectral shape. This suggests that they are due to two different but probably linked mechanisms.\\
The similarity between V404 Cyg and Cyg X-1  spectral shapes in the high energy domain suggests that the second component could be related to a jet contribution extending from radio up to hard X-rays.
The radio flaring activity reported during the outburst \citep{radioatel} could be used to test this hypothesis.
Polarization measurements could give a decisive answer. However, if the extreme variability of the source also affects the jet component, it could cancel  any coherent polarization signal.\\
In addition, such an exceptional event requires a careful study of the 500 keV energy domain.
 In the presented observations,
we did not detect any narrow feature at 511 keV, on the timescales of hours or days. This means that if 511 keV photons are produced, the flux
does not exceed $2 \times 10^{-4}$ \phs\ (2~$\sigma$) for a width of 4.5 keV and a total duration of 150 ks. 

The large amount of high  statistics data  from V404 Cyg opens a large domain of new studies: spectral variability
 over the minute timescale, evolution of the high energy component, timing studies in the 
  hard X-rays/soft gamma-ray domains, etc. This means that what we will learn from this peculiar event will  give us a new view of this family of objects.

\section{summary and conclusion}
The analysis of the V404 Cyg 2015 outburst is still in its early stages. The multiwavelength
campaign triggered by the first observations ensures a huge amount of data. 
The hard X-ray domain enlightens the innermost regions of the source and  reveals the behavior of the most energetic particles at
work during the flaring activity. 
The source emission is satisfactorily described by a hot (40-50 keV) Comptonized component, 
with no need for a second component ('quiet state'), or by a  cooler Comptonizing plasma 
  (25-30 keV) plus a second component, possibly related to jet synchrotron emission, which appears above the thermal cutoff ('active state').\\
A search for a narrow  emission at 511 keV remained unsuccessful, but deeper studies are required to investigate the presence of any broad and/or shifted feature potentially related to the annihilation process.\\

\textit{INTEGRAL} has observed this source for 4 weeks. This will allow the  polarization to be investigated.
Also, information coming from all wavelengths will have to be correlated to reveal a more complete picture of the source and  
understand the physics at work.  In conclusion, V404 Cyg offers us, with this exceptional outburst, a unique opportunity 
to make a significant  breakthrough in  both the X-ray transient phenomenon and X-ray binary sytems.\\

\textit{Note: Results on the two next revolutions are reported in \citet{JeromeV404}, with a different spectral description. All the SPI spectra presented in our paper are made available in a fits format on the site given in footnote 2, in order to allow anybody to test different models.}
 
\section*{Acknowledgments}  The \textit{INTEGRAL} SPI project has been completed 
under the responsibility and leadership of CNES. The Italian co-authors acknowledge the Italian Space
Agency (ASI) for financial support under ASI/INAF agreement n. 2013-025-R.0.
   We are grateful to ASI, CEA, CNES, DLR, ESA, INTA, NASA and OSTC for support.
 

\begin{deluxetable*}{lcccccccc}
\tablewidth{0pt}
\tablecaption{Best fit parameters for individual spectra (Comptt + cutoff power law model)}
\tabletypesize{\scriptsize}
\tablehead{
\colhead{Spectrum label }
&\colhead{kT} 
&\colhead{$\tau$}
&\colhead{ $F_{[25-100 keV]}^*$ }
 &\colhead{Ecut}
&\colhead{$N_{[1 keV]}^{**}$}
&\colhead{$\chi_{red}^{2} $}\\
 &\colhead{keV}
&&\colhead{$10^{-2}~ph/cm^{2}~s$}
 &\colhead{keV}
&\colhead{$ph/cm^{2}~s~keV$}
&\colhead{(dof)}
 }
\startdata 
 
 Averaged spectrum          &  26.4 $\pm$ 2          & 1.6 $\pm$0.2    &   26.0 $\pm$ 0.9   &  241 $\pm$  40 & 1.4 $\pm$ 0.02  & 1.1 (35) \\

First small peak [1] (scw 19-20)    &  42 $\pm$  5           &  1.1  +0.2      &   43  $\pm$ 2.4    &not required    &      & 1.1  (36)  \\
"Off-flare" [2] (scw 26-32)     &   53 $^{+ 44}_{- 12}$  & 1.2  $\pm$ 0.5  &   5.7 $\pm$ 1.2     &not required    &          &   1.2 (34)  \\ 
 
Plateau   [3] (scw 33-36) &   28 $\pm$  3        &1.6  $\pm$ 0.2     &  50.5  $\pm$ 4    &  296 $\pm$  80  &   1.7     & 1.4 (35)  \\

first Maximum  (scw 37) [4] &   21 $\pm$  2       & 1.9  $\pm$ 0.2     &  174 $\pm$ 6   &  189  $^{+ 74}_{- 36}$   &   8 $\pm$ 1.6    & 1.23 (35) \\

2nd Maximum  (scw 43) [5] &  23 $\pm$  2       & 1.7  $\pm$ 0.2     &  156  $\pm$  7    &  189 $\pm$  40  &  5.4 $\pm$ 1.8 & 1.0 (35)  \\
  &&==&  common fit&  ==&&&\\
Averaged spectrum           & 27.8 $\pm$ 1         & 1.55  $\pm$ 0.1   &  25.9 $\pm$ 0.3   &  242 $\pm$  20 & 1.3 $\pm$ 0.03 & 1.2  (37)  \\
First small peak [1] (scw 19-20)     & "                    & "                 &  42.5  $\pm$ 1.2  &  300 (fix)     & 1.9 $\pm$ 0.1  & 1.1  (38) \\
"Off-flare" [2]   (scw 26-32)      &   43  $\pm$ 4      & "                 &  4.9 $\pm$  1.3  &    "           & 0.2 $\pm$ 0.1 &  1.2  (34)  \\
Plateau    [3] (scw 33-36) &  idem [1]           &   "               &  50  $\pm$ 1.3   & "             &  1.8 $\pm$ 0.07 & 1.3 (38) \\
Maximum  (scw 37) [4]        &   25 $\pm$  1       & "                 &  183.8 $\pm$  1.8    &  "         &  4.3 $\pm$ 0.15 &  1.3  (38) \\
2nd Maximum  (scw 43) [5]  &      "             &  "                &  164.4  $\pm$  1.9 & "           &  2.6 $\pm$ 0.16 &1.4  (38)\\
 && == & 100s timescale&  ==&&\\
100s Max bin  [6]             & 22.5 $\pm$  1       & 1.55 (fix) &   680 $\pm$  12 & 300 (fix)&  9.8 $\pm$ 1 & 0.93  (36)\\

\enddata
\tablecomments{Parameters obtained for the individual spectra displayed in Fig.\ref{fig:6spectres}.
Top: Each spectrum is fit individually. Bottom: Spectra are fit simultaneously, with some parameters forced to be equal, to have the minimum of free parameters. The $kT_{0}$ is fixed to 7 keV, except for [1] which requires a value of 6.5 keV. Photon index of the cutoff power law component is fixed to 1.6.\\
 $^{*}$ For the Comptonization component $^{**}$ Cutoff powerlaw normalisation at 1 keV. 
}
\label{tab:fitindiv} 
\end{deluxetable*}

\end{document}